\documentclass[aps,pra,twocolumn]{revtex4-1}

\usepackage{graphicx}
\usepackage{amsmath}
\usepackage{amssymb}
\usepackage{hyperref}
\usepackage[utf8]{inputenc}
\usepackage{mathtools}
\usepackage[english]{babel}
\usepackage{lipsum}
\usepackage{bbm}
\hypersetup{colorlinks=true, linkcolor=blue, citecolor=blue, urlcolor=blue}
\usepackage{xcolor}
\usepackage{braket}

\newcommand{\ie}{i.\,e.,\ }

\newcommand{\cf}{cf.\ }
\newcommand{\re}{\mathrm{Re}}

\newcommand{\fref}[1]{\text{Fig.}~\ref{#1}}
\newcommand{\ffref}[1]{\text{Figs.}~\ref{#1}}
\newcommand{\eref}[1]{\text{Eq.}~\eqref{#1}}
\newcommand{\eeref}[1]{\text{Eqs.}~\eqref{#1}}

\begin{document}
\title{Many-Body Phases of a Planar Bose-Einstein Condensate with Cavity-Induced Spin-Orbit Coupling}
\author{Stefan Ostermann}
\email{stefan.ostermann@uibk.ac.at}
\affiliation{Institut f\"ur Theoretische Physik, Universit\"at Innsbruck, Technikerstra{\ss}e~21a, A-6020~Innsbruck, Austria}
\author{Helmut Ritsch}
\affiliation{Institut f\"ur Theoretische Physik, Universit\"at Innsbruck, Technikerstra{\ss}e~21a, A-6020~Innsbruck, Austria}
\author{Farokh Mivehvar}
\email{farokh.mivehvar@uibk.ac.at}
\affiliation{Institut f\"ur Theoretische Physik, Universit\"at Innsbruck, Technikerstra{\ss}e~21a, A-6020~Innsbruck, Austria}

\begin{abstract}
We explore the many-body phases of a two-dimensional Bose-Einstein condensate with cavity-mediated dynamic spin-orbit coupling. By the help of two transverse non-interfering, counterpropagating pump lasers and a single standing-wave cavity mode, two degenerate Zeeman sub-levels of the quantum gas are Raman coupled in a double-$\Lambda$-configuration. Beyond a critical pump strength the cavity mode is populated via coherent superradiant Raman scattering from the two pump lasers, leading to the appearance of a dynamical spin-orbit coupling for the atoms. We identify three quantum phases with distinct atomic and photonic properties:  the normal ``homogeneous'' phase, the superradiant ``spin-helix'' phase, and the superradiant ``supersolid spin-density-wave'' phase. The latter exhibits  an emergent periodic atomic density distribution with an orthorhombic centered rectangular-lattice structure due to the interplay between the coherent photon scattering into the resonator and the collision-induced momentum coupling. The transverse lattice spacing of the emergent  crystal is set by the dynamic spin-orbit coupling.
\end{abstract}

\maketitle

\section{Introduction}

A Bose-Einstein condensate (BEC) is a phase of quantum matter with intriguing properties and no classical counterpart~\cite{davis_bose-einstein_1995,anderson_observation_1995,pitaevskii_bose-einstein_2016}. The controllablility and tunability of atomic BECs has seen an immense development over the past decades such that BECs have become fundamental environments for studying many-body effects in the quantum regime~\cite{lewenstein_ultracold_2007,bloch_many-body_2008,gross_quantum_2017}. A particularly fascinating research area is the study of spin-orbit-coupled BECs both in free space~\cite{wang_spin-orbit_2010,campbell_realistic_2011,lin_spinorbit-coupled_2011, martone_anisotropic_2012,li_quantum_2012,galitski_spinorbit_2013,ji_softening_2015,zhang_properties_2016,aidelsburger_cold_2016,li_stripe_2017,putra_spatial_2020} and in optical lattices~\cite{beugeling_topological_2012,radic_exotic_2012,cole_bose-hubbard_2012, kennedy_spin-orbit_2013, cai2012magnetic, piraud2014quantum,xu2014mott,wu_realization_2016, yamamoto_quantum_2017, li_topological_2018, li_spin_2019, kartashov_stable_2019}. Spin-orbit-coupled BECs have the potential for investigating complex phases of quantum matter, such as topological states~\cite{Cooper2019Topological}, that go beyond traditional condensed matter physics.

Recently BECs coupled to dynamic electromagnetic fields of optical resonators have been established as a promising platform for investigating collective self-organizing phenomena in quantum regimes under well-controlled conditions~\cite{ritsch_cold_2013}. The cavity-enhanced back-action of the BEC on the cavity light fields and vice versa creates dynamic optical potentials and long-range atom-atom interactions giving rise to intriguing collective phases~\cite{nagorny_collective_2003,brennecke_cavity_2007,baumann_dicke_2010,klinder_observation_2015,landig_quantum_2016}. Recent milestones in the field of many-body cavity QED with BECs include the experimental and/or proposed realization of intriguing nonequilibrium effects and quantum phases such as supersolids~\cite{leonard_supersolid_2017, mivehvar_driven-dissipative_2018, schuster_supersolid_2020}, non-trivial spin orders~\cite{landini_formation_2018, kroeze_spinor_2018, mivehvar_cavity-quantum-electrodynamical_2019, colella_antiferromagnetic_2019,ostermann_cavity-induced_2019, bentsen_integrable_2019}, dynamical synthetic spin-orbit coupling~\cite{mivehvar_synthetic_2014,dong_cavity-assisted_2014,padhi_spin-orbit-coupled_2014,deng_bose-einstein_2014,mivehvar_enhanced_2015, Catalin-Mihai2019Cavity, kroeze_dynamical_2019}, and emergent quasicrystalline symmetries~\cite{mivehvar_emergent_2019}. For most of the self-ordering phenomena the coherent resonator fields play a decisive role in the self-organization process, while the coherence properties of the condensate are not substantial in the formation of periodic density modulations. Hence, typically self-organization for dilute atomic gases in optical resonators can also be studied with cold thermal atomic clouds~\cite{domokos_collective_2002,black_observation_2003}. This implies that, in general, the emergent density modulation exhibits a periodicity set via the cavity resonance wavelength.

\begin{figure}[b!]
\centering
\includegraphics[width=0.49\textwidth]{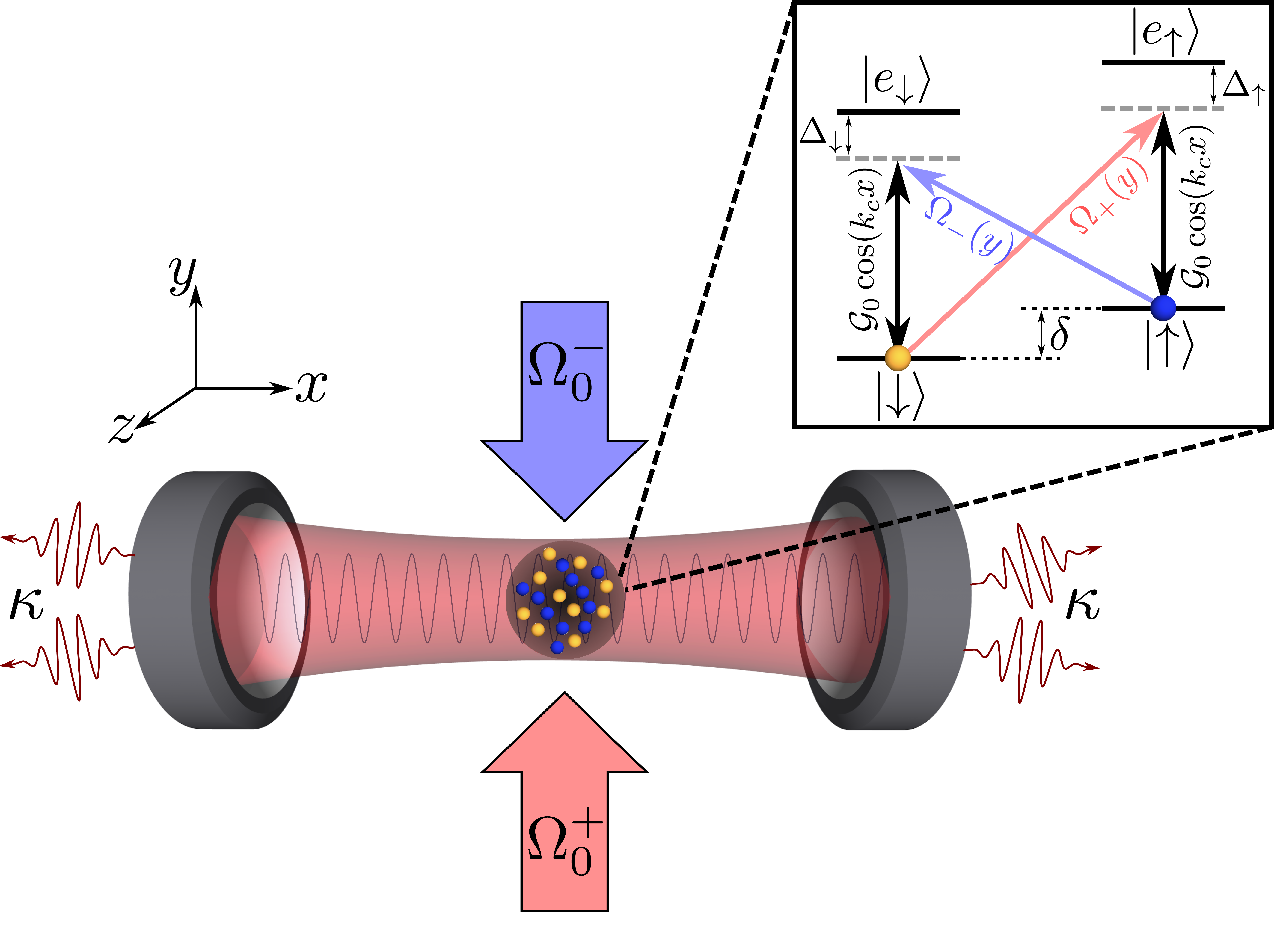}
\caption{Sketch of the system.  A two-component BEC in 2D couples strongly to a single $\pi$ mode of a standing-wave resonator with the maximum  vacuum Rabi frequency $\mathcal{G}_0$. Two counterpropagating $\sigma^\pm$ running-wave pump lasers with Rabi rates $\Omega_0^{\pm}$ illuminate the atoms from the transverse direction. The microscopic atom-photon coupling scheme is shown in the inset.}
\label{fig:setup}
\end{figure}

In this work we show that for an interacting two-component BEC in two dimensions (2D) strongly coupled to a single standing-wave mode of a cavity new phenomena beyond this paradigm can appear. The studied setup builds upon the recent theoretical proposal for cavity-induced spin self-ordering~\cite{mivehvar_cavity-quantum-electrodynamical_2019} and the recent experimental realization of a dynamic spin-orbit coupling~\cite{kroeze_dynamical_2019} in non-interacting BECs inside cavities, by incorporating the two-body contact interactions and exploring their interplay with coherent photon scattering. In particular, dynamical spin-orbit coupling is induced in our setup for the interacting two-component BEC via two counterpropagating pump lasers and the standing-wave cavity mode as shown schematically in Fig.~\ref{fig:setup}. We find that the presence of the cavity has a significant impact on the single-particle and many-body physics. Most strikingly we identify a parameter regime where the many-body ground-state density distribution spontaneously forms a centered orthorhombic crystal. This density pattern emerges as a consequence of two combined effects: coherent superradiant scattering of laser photons into the cavity and matter-wave interference of the BEC momentum components due to elastic collisions in 2D. Since the off-diagonal long-range order and interactions of the BEC simultaneously play a crucial role in the formation of this self-ordered pattern, our results have the potential to establish a new paradigm in the self-ordering of BECs in resonators. The crystalline state possesses supersolid characteristics along the transverse pump direction because it breaks the continuous translational symmetry in $y$ direction. This phase is intimately related to the supersolid stripe phase in 1D spin-orbit-coupled BECs in free space. However, in contrast to free space, we find that the spin-orbit coupling and the supersolid phase persist for a wider parameter regime. Furthermore, the density modulation in the supersolid phase is much more pronounced and multiple momentum components are populated, which should facilitate the (destructive or non-destructive) experimental detection of the supersolid phase.

The paper is organized as follows. First we introduce the theoretical model in Section~\ref{sec:theo_model}. In Section~\ref{sec:phase_diag} we discuss the numerically obtained phase diagram. We divide the discussion into two parts; first we focus on non-interacting atoms and show that the presence of the cavity already changes the physics of the system on this level. We then show that including two-body interactions leads to the appearance of an additional phase---the ``supersolid spin-density-wave'' phase. The interplay of coherent photon scattering and collision-induced momentum coupling in this phase is then discussed. To clarify this point, in Section~\ref{sec:groundstates} we illustrate exemplary ground states for the different phases and provide intuitive explanations for the different phases based on the many-body Hamiltonian in momentum space. We conclude and give an outlook on future perspectives in Section~\ref{sec:conclusion}.

\section{Theoretical Model}\label{sec:theo_model}
We consider a multi-component BEC in 2D (the $x$-$y$ plane) placed inside a standing-wave optical cavity; see~\fref{fig:setup}. The bosonic atoms are assumed to have four levels with two pseudospin ground states $\ket{\downarrow}$ and $\ket{\uparrow}$ and the corresponding excited states $\ket{e_{\downarrow, \uparrow}}$. The atoms couple to a single resonator mode with resonance frequency $\omega_c$ via the transitions $\ket{\downarrow}\leftrightarrow\ket{e_\downarrow}$ and $\ket{\uparrow}\leftrightarrow\ket{e_\uparrow}$ with a coupling strength $\mathcal{G}(x,y)=\mathcal{G}_0 \cos(k_c x)$. $k_c=2\pi/\lambda_c=\omega_c/c$ is the cavity mode wavenumber related to the cavity wavelenth $\lambda_c$, with $c$ being the speed of light. In addition, two counterpropagating running-wave lasers with frequencies $\omega_{p_+}$ and $\omega_{p_-}$ illuminate the atoms perpendicular to the cavity axis ($y$ direction, \cf~\fref{fig:setup}). These additional pump lasers drive the transitions $\ket{\downarrow} \leftrightarrow \ket{e_\uparrow}$ and $\ket{\uparrow} \leftrightarrow \ket{e_\downarrow}$ with position dependent Rabi rates $\Omega_\pm (\mathbf{r})= \Omega_{0}^\pm e^{\pm i k_c y}$. Assuming large detuning of the pump and cavity frequencies from any atomic resonances ($\Delta_{\downarrow,\uparrow} \gg 1$) allows the adiabatic elimination of the atomic excited states $\ket{e_{\downarrow,\uparrow}}$~\cite{mivehvar_cavity-quantum-electrodynamical_2019}. In this case the system resembles a spin-1/2 configuration, where the pseudospin is coupled to the cavity mode via two-photon Raman transitions.

For the remainder of this work we focus on the case where the two pseudospin states are degenerate,~\ie $\delta=0$, where $\delta$ is the energy difference between the two states $\ket{\uparrow}$ and $\ket{\downarrow}$. We also choose $\omega_{p_+}=\omega_{p_-}\equiv\omega_p$  and $\Delta_\downarrow=\Delta_\uparrow\equiv\Delta_{a}$. In addition, we restrict our analysis to balanced pump intensities $\Omega_0^+=\Omega_0^-\equiv\Omega_0$. Under these conditions the single-particle spinor Hamiltonian in the rotating-wave approximation is given in the matrix form by
\begin{equation}
\hat{h}=
\begin{bmatrix}
\frac{\hat{\mathbf{p}}^2}{2m} + \hat{a}^\dagger \hat{a} \, U(\mathbf{r})
& (\hat{a}^\dagger+\hat{a})\,\Omega_R(\mathbf{r}) \\
(\hat{a}^\dagger+\hat{a})\,\Omega_R^*(\mathbf{r})
&\frac{\hat{\mathbf{p}}^2}{2m} +\hat{a}^\dagger \hat{a} \, U(\mathbf{r})
\end{bmatrix},
\label{eqn:Ham_SP}
\end{equation}
where we have defined the functions $U(\mathbf{r}) = \hbar U_0  \cos^2(k_c x)$ and $\Omega_R(\mathbf{r}) = \hbar \eta_0\cos(k_c x)e^{i k_c y}$, $\hat{\mathbf{p}}=-i\hbar\nabla$ is the canonical momentum operator, and $\hat{a}$ ($\hat{a}^\dagger$) is the bosonic photon annihilation (creation) operators of the cavity mode. We have also introduced the maximum potential energy per photon $\hbar U_0 \coloneqq \hbar |\mathcal{G}_0|^2/\Delta_a$ and the maximum effective cavity pump strength $\eta_0\coloneqq \mathcal{G}_0 \Omega_0^*/\Delta_a$. 

The Hamiltonian in~\eref{eqn:Ham_SP} exhibits the typical spin-orbit coupling nature,~\ie different pseudospin components couple to different momenta, by equal contributions of Rashba~\cite{bychkov_oscillatory_1984} and Dresselhaus~\cite{dresselhaus_spin-orbit_1955} couplings. Hence, the employed Raman transitions give rise to a spin-orbit coupling similar to those found in solid state materials where the linear crystal momentum interacts with the spin of an electron~\cite{ashcroft_solid_1976}. This also implies that the canonical momentum $\hat{\mathbf{p}}=-i \hbar \nabla$ no longer coincides with the kinetic momenta~\cite{pitaevskii_bose-einstein_2016,kroeze_dynamical_2019}
\begin{align} \label{eqn:physical_mom}
\hat{\mathbf{P}}^\uparrow&=\hat{\mathbf{p}}+\frac{1}{2}\hbar k_c\mathbf{e}_y,\nonumber\\
\hat{\mathbf{P}}^\downarrow&=\hat{\mathbf{p}}-\frac{1}{2}\hbar k_c\mathbf{e}_y,
\end{align}
of the two pseudospin components ($\mathbf{e}_y$ is the unit vector along the $y$ direction). Note that spin-orbit coupling only occurs along the direction of the pump lasers,~\ie the $y$ direction. In this direction the Hamiltonian~\eqref{eqn:Ham_SP} exhibits a continuous screwlike symmetry, \ie the Hamiltonian is invariant under the unitary transformation $\mathcal{U}=e^{i\Delta y(\hat{p}_y+\hbar k_c\hat{\sigma}_z/2)/\hbar}$. This corresponds to a combination of a rigid translation along the $y$ axis by a distance $\Delta y$ and a simultaneous spin rotation by an angle $k_c\Delta y$ around the $z$ axis.

Due to the additional coupling to the cavity, spin-orbit coupling only occurs if the cavity mode is populated, \ie $\langle\hat{a}\rangle\neq 0$. Hence, the Raman transitions are only driven if the effective pump strength $\eta_0$ exceeds the critical value~\cite{mivehvar_cavity-quantum-electrodynamical_2019},
\begin{equation}
\sqrt{N}\eta_c=\sqrt{-\frac{(\Delta_c-NU_0)^2+\kappa^2}{\Delta_c-N U_0}}\omega_\mathrm{rec},
\label{eqn:eta_crit}
\end{equation}
where $\omega_\mathrm{rec}\coloneqq\hbar k_c^2/(2m)$ is the recoil frequency, $\Delta_c\coloneqq\omega_p-\omega_c$ is the cavity detuning with respect to the pump lasers, $\kappa$ denotes the cavity decay rate, and $N$ is the total particle number.

The many-body Hamiltonian, plus the free cavity Hamiltonian, in the second quantized formalism reads
\begin{equation}
\hat{H}=\int d\mathbf{r}\hat{\Psi}^\dagger \hat{h} \hat{\Psi} - \hbar \Delta_c \hat{a}^\dagger \hat{a} + \hat{H}_\mathrm{int},
\label{eqn:Ham_MB_x}
\end{equation}
with the spinor $\hat{\Psi}(\mathbf{r})=[\hat{\Psi}_\uparrow(\mathbf{r}),\hat{\Psi}_\downarrow(\mathbf{r})]^\top$ consisting of the single component bosonic field operators $\hat{\Psi}_{\uparrow,\downarrow}(\mathbf{r})$. The interaction Hamiltonian,
\begin{align}
\hat{H}_\mathrm{int}&=\frac{g}{2} \sum_ {\tau\in\{\uparrow,\downarrow\}}\int  
\hat{\Psi}_\tau^\dagger(\mathbf{r})\hat{\Psi}_\tau^\dagger(\mathbf{r}) \hat{\Psi}_\tau(\mathbf{r})\hat{\Psi}_\tau(\mathbf{r}) d\mathbf{r}
\nonumber\\
&+g_{\uparrow\downarrow} \int 
\hat{\Psi}_{\uparrow}^\dagger(\mathbf{r})\hat{\Psi}_{\downarrow}^\dagger(\mathbf{r}) \hat{\Psi}_\downarrow(\mathbf{r})\hat{\Psi}_{\uparrow}(\mathbf{r})
d\mathbf{r},
\label{eqn:Ham_MB_2_body_int}
\end{align}
takes into account elastic two-body contact interactions between atoms of the same pseudospin with the interaction strength $g$ and opposite pseudospins with the interaction strength $g_{\uparrow\downarrow}$.

\begin{figure}
\centering
\includegraphics[width=0.46\textwidth]{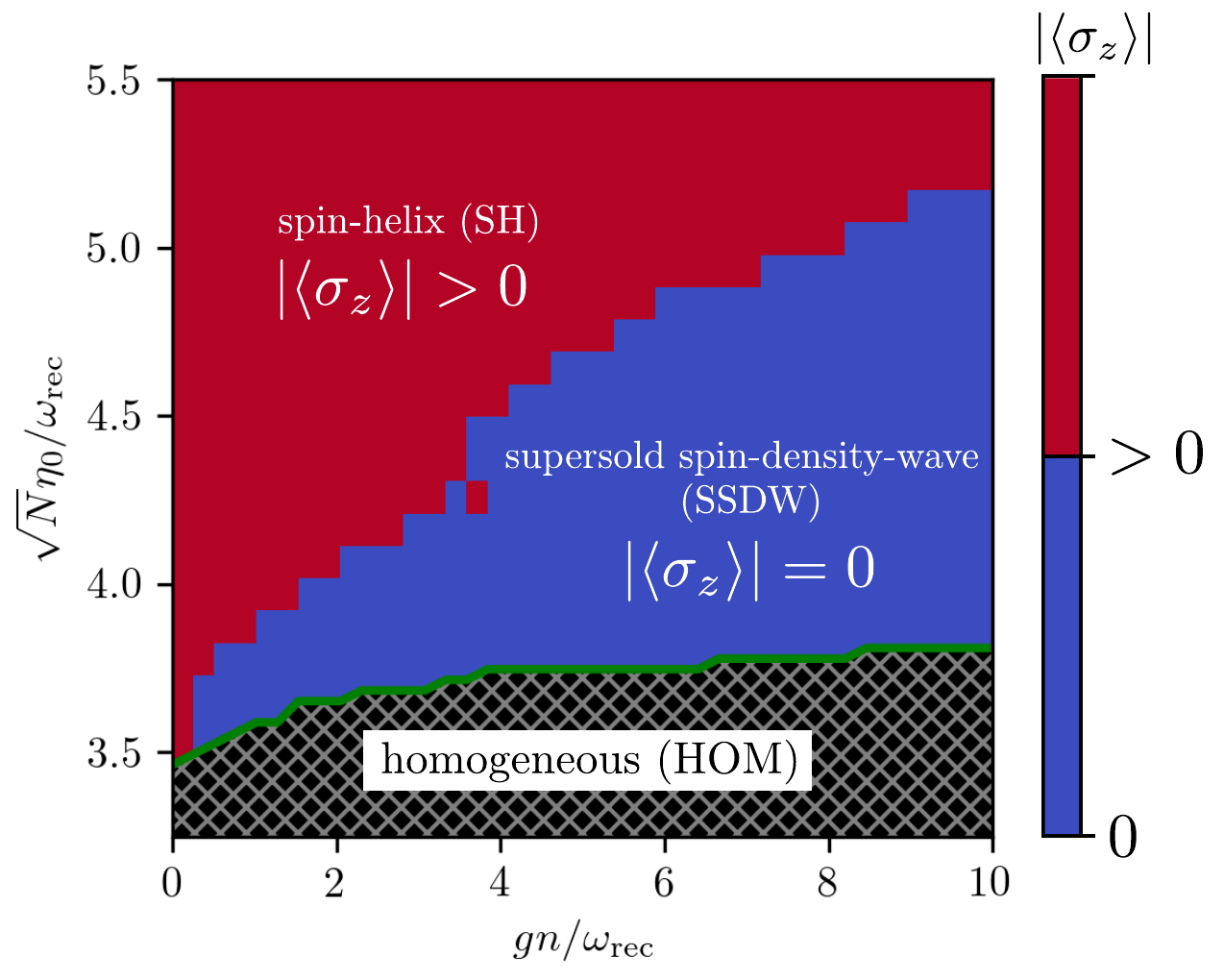} 
\caption{Many-body phase diagram for $\gamma=0.8$ as a function of effective pump strength $\sqrt{N}\eta_0$ and interaction strength $gn$ ($n=N/V$ is the total particle density). The black white hatched region indicates the parameter regime where the density is homogeneous (HOM) and the cavity-mode amplitude is zero $|\alpha|=0$. In the red region where $|\langle\sigma_z\rangle| > 0$ the spin-helix (SH) phase is the ground state. In the blue region the spin imbalance vanishes $|\langle\sigma_z\rangle|=0$ and a supersolid spin-density-wave (SSDW) phase is realized. The other parameters are chosen as $(\Delta_c,N U_0,\kappa)=(-10,-1,5)\omega_\mathrm{rec}$.}
\label{fig:MB_phase_diag}
\end{figure}

The system's dynamics is governed by the Heisenberg equations of motion for the atomic and photonic field operators $\hat{\Psi}_\tau(\mathbf{r},t)$ and $\hat{a}(t)$ (in the Heisenberg representation): $i\hbar\partial_t \hat{\Psi}_\tau (\mathbf{r},t) =[\hat{\Psi}_\tau(\mathbf{r},t),\hat{H}]$ with $\tau\in\{\uparrow,\downarrow\}$ and $i\hbar\partial_t \hat{a}=[\hat{a},\hat{H}]-i\kappa\hat{a}$. The non-Hermitian term $\propto - i \kappa$ in the equation of motion for the cavity mode accounts for the loss of cavity photons at a rate $2\kappa$.
In the mean-field approximation, where the field operators are replaced by their expectation values $\hat{\Psi}_\tau(\mathbf{r},t) \rightarrow \langle \hat{\Psi}_\tau(\mathbf{r},t) \rangle \equiv \psi_\tau(\mathbf{r},t)$ and  $\hat{a} \rightarrow \langle \hat{a} \rangle\equiv\alpha$, this results in three coupled equations for the BEC mean-field wave-functions $\psi_{\uparrow,\downarrow}$ and the cavity-field coherent amplitude $\alpha$, respectively,
\begin{subequations}
\begin{align}
i\hbar\partial_t\psi_\uparrow&=\Big[-\frac{\hbar^2\nabla^2}{2m}+|\alpha|^2 U(\mathbf{r})
+g|\psi_\uparrow|^2 + g_{\uparrow\downarrow}|\psi_\downarrow|^2\Big]\psi_\uparrow \nonumber\\
&\quad \quad + 2\re(\alpha) \Omega_R(\mathbf{r})\psi_\downarrow,
\label{eqn:coupled_GPE_a}\\
i\hbar\partial_t\psi_\downarrow&=\Big[-\frac{\hbar^2\nabla^2}{2m} + |\alpha|^2 U(\mathbf{r})
+ g|\psi_\downarrow|^2 + g_{\uparrow\downarrow}|\psi_\uparrow|^2\Big]\psi_\downarrow \nonumber\\
&\quad \quad + 2\re(\alpha)\Omega_R^*(\mathbf{r})\psi_\uparrow,
\label{eqn:coupled_GPE_b}\\
i\hbar\partial_t\alpha&=\hbar\left[-\Delta_c-i\kappa + U_0 \mathcal{B}\right]\alpha
+\hbar\eta_0 \re(\alpha)\mathcal{S}.
\label{eqn:mode_dyn}
\end{align}
\label{eqn:dynamics}%
\end{subequations}
In the last line we introduced the bunching parameter $\mathcal{B}\coloneqq\int d\mathbf{r}\cos^2(k_c x) \left(|\psi_\uparrow|^2 + |\psi_\downarrow|^2\right)$ and the spin order parameter  $\mathcal{S}\coloneqq\int d\mathbf{r} \cos(k_c x)\left(e^{i k_c y}\psi^*_\uparrow\psi_\downarrow + e^{-i k_c y}\psi^*_\downarrow\psi_\uparrow\right)$, characterizing the superradiant phase transition. The total particle number is fixed via $N=\int d\mathbf{r}(|\psi_\uparrow|^2 + |\psi_\downarrow|^2)$. This set of equations again exhibits the dynamic nature of spin-orbit coupling due to the presence of the cavity field. Since the Raman coupling terms [last terms in~\eqref{eqn:coupled_GPE_a} and~\eqref{eqn:coupled_GPE_b}] explicitly depend on the value of $\alpha$, the spin-orbit coupling depends on the non-linear cavity-field dynamics governed by Eq.~\eqref{eqn:mode_dyn}. In addition, there is \emph{no} spin-orbit coupling below the superradiant self-ordering phase transition. However, once the cavity mode is populated,~\ie $\alpha \neq 0$, via two-photon Raman processes, the spin-orbit coupling for the BEC sets in. In~\eref{eqn:mode_dyn} the pump term $\propto\eta_0$ only depends on the spin order parameter $\mathcal{S}$, which shows that the cavity mode can only be populated by spin changing Raman processes, but not via scattering from the BEC density~\cite{deng_bose-einstein_2014, masalaeva2020spin}. In this respect the system differs fundamentally from common quantum-gas--cavity systems exhibiting a superradiant self-ordering phase transition due to photon scattering from the atomic density distribution~\cite{ritsch_cold_2013}.

\section{Phase diagram}\label{sec:phase_diag}

\begin{figure}
\centering
\includegraphics[width=0.48\textwidth]{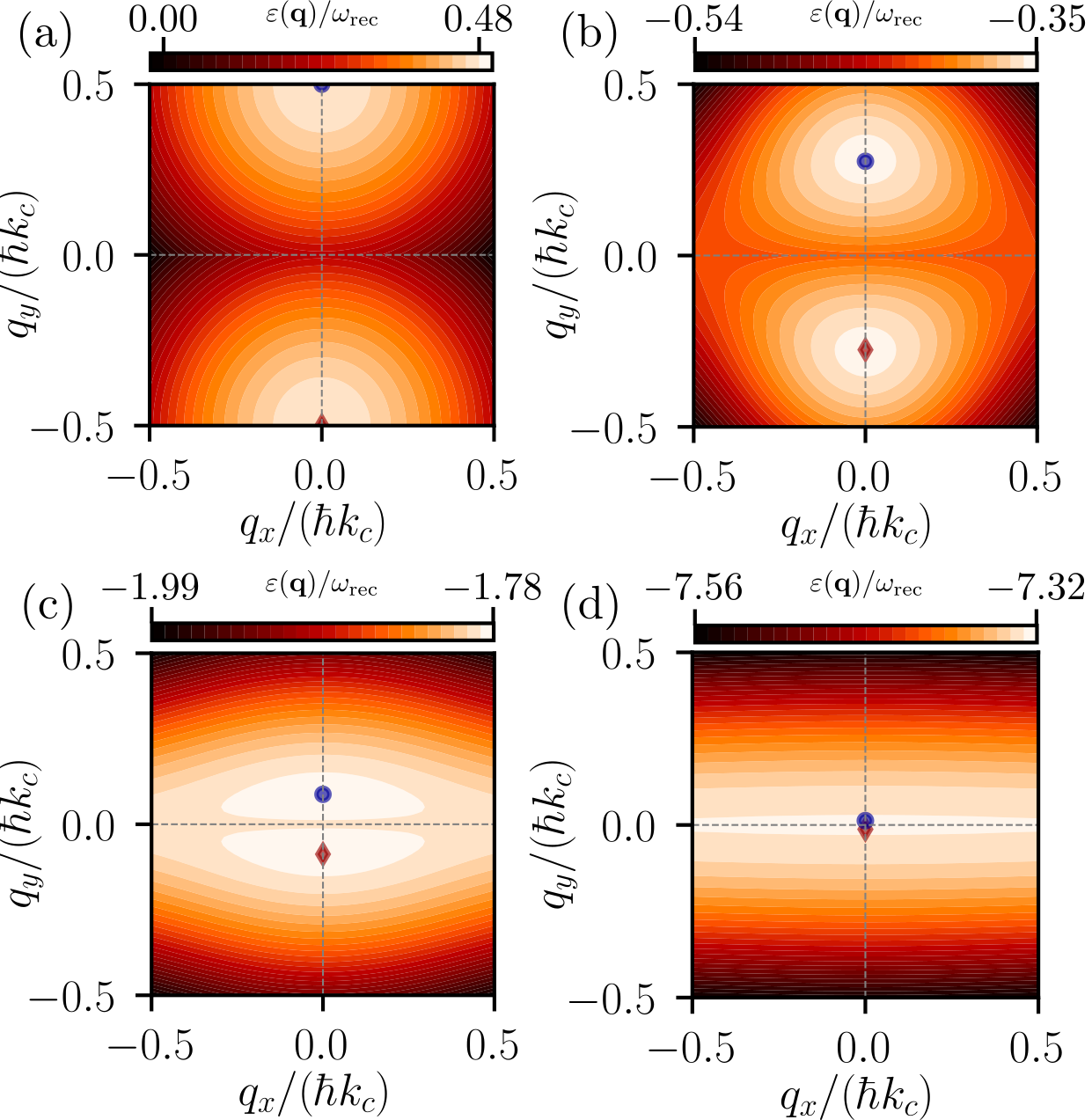} 
\caption{Exemplary contour plots of the single-particle energy dispersion for $\sqrt{N}\eta_0=(1.0,3.8,5.0,8.0)\omega_\mathrm{rec}$, respectively (a)--(d). The blue dot and the red diamond mark the two minima of the dispersion. The functional dependence of these minima on $\eta_0$ is illustrated in~\fref{fig:PD_non-int}. All other parameters are the same as in~\fref{fig:MB_phase_diag}.}
\label{fig:dispersion}
\end{figure}

We numerically calculate the steady state of the system for a fixed parameter set by applying a self-consistent algorithm. For a given initial (random) value of the cavity-field amplitude $\alpha$ we find the atomic ground states $\psi_{\downarrow,\uparrow}$. With this atomic ground states we then calculate the corresponding steady-state cavity-field amplitude $\alpha_\mathrm{ss}=\eta_0\mathcal{S}/(\Delta_c+i \kappa - U_0\mathcal{B})$, obtained by applying the steady-state condition $\partial_t\alpha=0$ in~\eref{eqn:mode_dyn}. This steady-state field amplitude is again plugged back into the atomic GP equations. The self-consistent loop continues until the convergence.

The full phase diagram of the system in the parameter plane of the pump strength $\sqrt{N}\eta_0$ versus the intra-species two-body interaction strength $gn$ is illustrated in Fig.~\ref{fig:MB_phase_diag}. It is instructive to analyze the phase diagram in two steps. First the generic ground-state properties for the non-interacting case (cut for $g=g_{\uparrow\downarrow}=0$) is presented in the following and afterwards the effect of interactions leading to the full phase diagram is discussed. A more detailed discussion of each phase is presented in Section~\ref{sec:groundstates}.  The different phases can be characterized and distinguished via three global quantities: the total spin imbalance  $\langle\sigma_z\rangle=\int d\mathbf{r}(|\psi_\uparrow|^2-|\psi_\downarrow|^2)$, the quasi-momentum at which the dispersion relation has its minimum, and the steady-state cavity-field amplitude $|\alpha_\mathrm{ss}|$. The first two quantities are standard quantities to characterize the different phases of spin-orbit-coupled BECs~\cite{zhang_properties_2016}, whereas the cavity-mode amplitude characterizes the superradiant phase(s) of the system. 

\subsection{Non-interacting BEC}
\label{sec:ideal-BEC}

\begin{figure}
\centering
\includegraphics[width=0.42\textwidth]{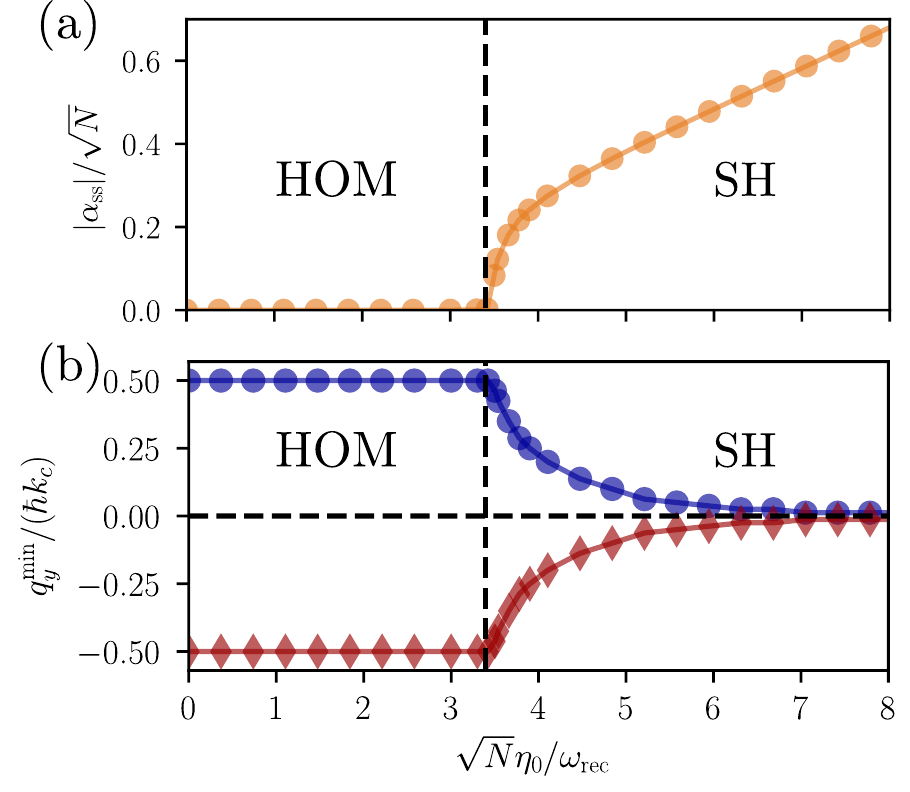}
\caption{Superradiant phase transition and the emergence of a dynamical spin-orbit coupling in the single-particle regime. (a) Re-scaled steady-state cavity-field amplitude (orange circles) as a function of the pump strength. The numerically obtained critical pump strength is $\sqrt{N}\eta_c=3.43\omega_\mathrm{rec}$. (b) The $y$-components of the quasi-momenta $q_y^\mathrm{min}$ at which the dispersion relation has its minima as a function of the pump strength. Blue circles (red diamonds) indicate the minima position for the positive (negative) quasi-momentum; see also~\fref{fig:dispersion}. The solid lines are a guide for the eye. All other parameters are the same as in~\fref{fig:MB_phase_diag}.}
\label{fig:PD_non-int}
\end{figure}

One fundamental implication of spin-orbit coupling is that the energy spectrum can possess several global minima at non-zero quasi-momenta. To find the quasi-momentum (or quasi-momenta) at which condensation takes place, we use a Bloch ansatz for the wave functions $\psi_{\uparrow,\downarrow}(\mathbf{r})=e^{i\mathbf{q}\cdot\mathbf{r}}u_{\uparrow,\downarrow}(\mathbf{r})$ where the quasi momentum $\mathbf{q}$ lies in the first Brillouin zone (BZ) $\mathbf{q} \in [-k/2 , k/2] \times [-k/2,k/2]$. We then find the self-consistent lowest energy state for the single-particle problem ($g=g_{\uparrow\downarrow}=0$) on a space $\mathbf{r}\in[0,\lambda] \times [0,\lambda]$ for all quasi-momenta in the first BZ. The corresponding ground state energies as a function of the quasi-momentum generate the energy dispersion for a given parameter set. Note that due to the non-linear coupling to the cavity mode described via the terms proportional to $U(\mathbf{r})$ and $\Omega_R(\mathbf{r})$ it is not possible to obtain a compact analytical formula for the dispersion relation. Exemplary single-particle energy dispersions for different effective pump strengths $\eta_0$ are shown in~\fref{fig:dispersion}. The dispersion exhibits two degenerate global minima, which is a hallmark of a spin-orbit coupling. The states corresponding to the global minima of the dispersion (indicated by blue dot for $q_y^\mathrm{min}>0$ and red diamond for $q_y^\mathrm{min}<0$ in~\fref{fig:dispersion}) are the possible ground states for the given parameter set.

In~\fref{fig:PD_non-int} the steady-state cavity-field amplitude and the $y$-components of the quasi-momenta at which the dispersion has its minimum $q_y^\mathrm{min}$ are shown as a function of the effective pump strength $\eta_0$. The energy dispersion exhibits the typical double-minima nature known from spin-orbit coupling in free space~\cite{lin_spinorbit-coupled_2011}. However, due to the additional coupling to the cavity, spin-orbit coupling only sets in beyond the critical pump strength. For pump intensities below the critical value the cavity-mode amplitude remains zero [see~\fref{fig:PD_non-int}(a)] and the dispersion relation possesses two minima at the edge of the BZ. If the intensity of the Raman beams exceeds the critical value given in~\eref{eqn:eta_crit}, the cavity mode is populated due to the second-order superradiant phase transition [see~\fref{fig:PD_non-int}(a)]. The numerically obtained value for the critical pump strength $\sqrt{N}\eta_c^\mathrm{num}=3.43\omega_\mathrm{rec}$ coincides (up to three significant digits) exactly with the analytical value $\sqrt{N}\eta_c=3.43\omega_\mathrm{rec}$ calculated from~\eref{eqn:eta_crit}. The non-vanishing cavity field amplitude implies that beyond the critical pump strength $\eta_c$ a dynamical spin-orbit coupling emerges~\cite{kroeze_dynamical_2019}. As a result, the single-particle energy dispersion exhibits two symmetric minima \emph{inside} the first Brillouin zone due to the spin-orbit coupling. The position of the minima changes along the $y$ direction with increasing pump strength [see blue dots and red diamonds in~\fref{fig:dispersion}(b)--(d) and~\fref{fig:PD_non-int}(b)]. Each of those minima is capable of hosting a BEC and the system chooses spontaneously in which of the two minima to condense, therefore breaking the degeneracy of the single-particle spectrum. 

Figure~\ref{fig:PD_non-int}(b) unveils another important effect of the dynamic cavity field on the single particle physics. In contrast to the free space case where the dependence of $q_y^\mathrm{min}$ as a function of the pump strength $\eta_0$ is always concave, the curve shown in~\fref{fig:PD_non-int} is convex, $\partial^2 q_y^\mathrm{min}/(\partial\eta_0^2)>0$. The reason for this change in curvature is the coupling to the cavity, which gives rise to the terms $\propto\cos^2(k_c x)$ and $\propto \cos(k_c x)$ in the Hamiltonian in~\eref{eqn:Ham_SP}. These terms modify the dispersion relation correspondingly, resulting in the change of curvature. The change in curvature has an important physical consequence. In the free space case the concave nature of the curve results in the fact that the quasi-momentum at which condensation takes place is zero for sufficiently strong Raman coupling. This results in the so-called zero momentum phase where the dispersion only exhibits a single minimum. In the cavity-induced spin-orbit coupling case studied here, in contrast, this single-minimum phase only occurs at $\eta_0 \rightarrow \infty$. That is, the double minima in the single-particle energy dispersion---the hallmark of spin-orbit coupling---persists for a much wider parameter regime. This also implies that the spin imbalance $\langle\sigma_z\rangle$ is always non-zero in the single particle regime. Another important feature, which can be seen from the energy dispersions in Fig.~\ref{fig:dispersion}, is that for increasing pump strength $\eta_0$ the bands become flatter in the $q_x$ direction. This results from the increasing depth of the emerging cavity potential which increases the effective mass in the $x$ direction. 

This analysis shows that in the single-particle regime the system exhibits two distinct phases which we call the normal homogeneous (HOM) phase and the superradiant spin-helix (SH) phase. In the first phase (HOM) the BEC is homogeneous in both $x$ and $y$ directions and the cavity-field amplitude is zero. No spin-orbit coupling occurs in this regime. In the latter case (SH) the cavity mode is populated and spin-orbit coupling emerges. Since the cavity mode is populated the density and spin textures are modulated in $x$ direction and due to the cavity-induced spin orbit coupling the spin exhibits a helix in $y$ direction without any density modulations. For a detailed discussion of the properties of this phase we refer to section~\ref{sec:SH-phase}.

\subsection{Interacting BEC}

\begin{figure}
\centering
\includegraphics[width=0.49\textwidth]{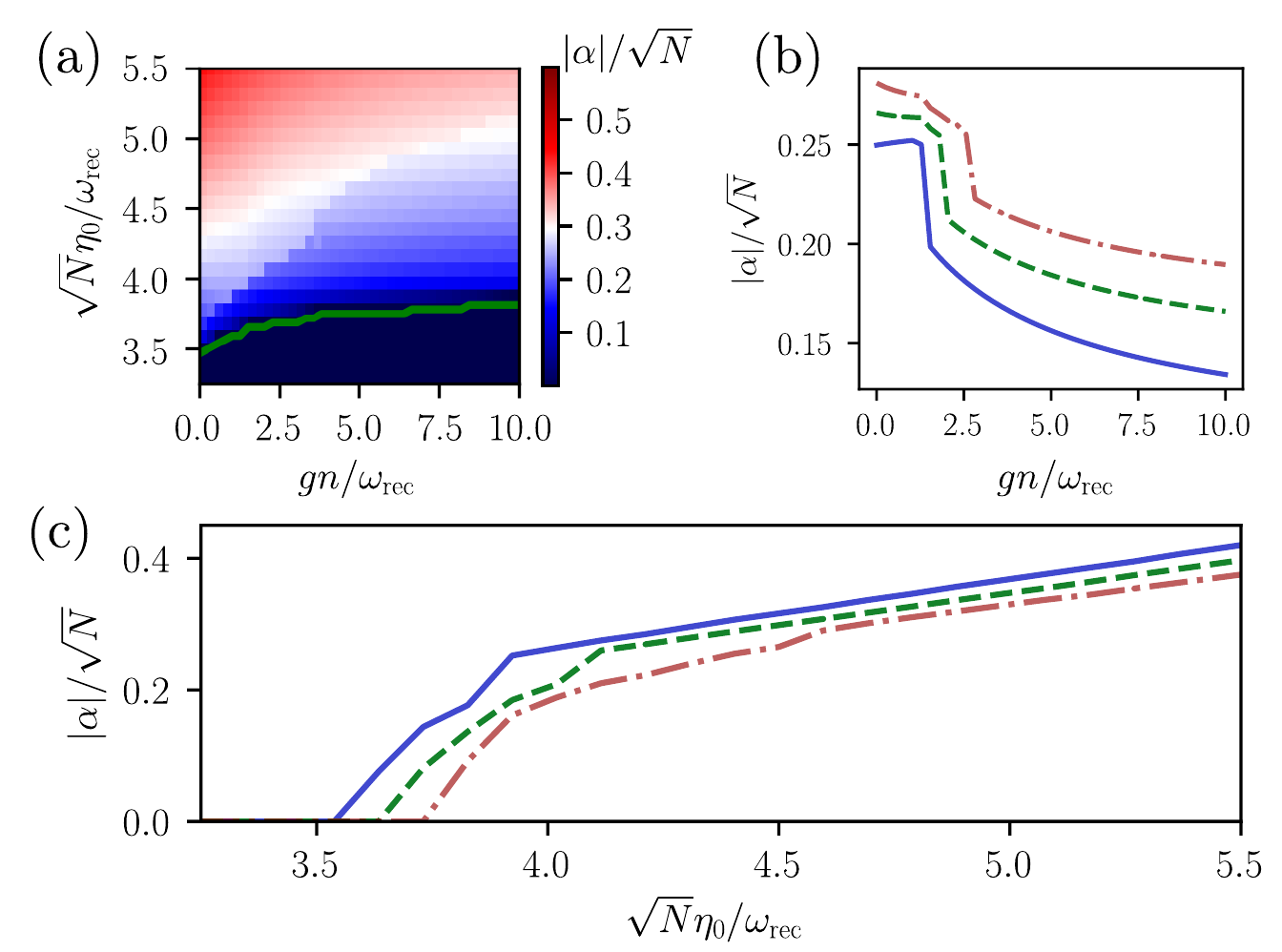} 
\caption{The absolute value of the cavity-field order parameter $|\alpha|$ (related to $\mathcal{S}$) in the same parameter space as the phase diagram in~\fref{fig:MB_phase_diag}. The quantum phase transitions can be monitored via probing the cavity mode amplitude $|\alpha|$. The panel (b) shows horizontal cuts of the field amplitude phase diagram shown in panel (a) for fixed $\sqrt{N}\eta_0=(3.92,4.02,4.11)\omega_\mathrm{rec}$ (solid blue, green dashed, red dash-dotted). Panel (c) shows vertical cuts for $gn=(3.63,4.12,4.88)\omega_\mathrm{rec}$. The cavity field amplitude exhibits a jump at the SH to SSDW transition. All other parameters are the same as in~\fref{fig:MB_phase_diag}.}
\label{fig:alpha_PD}
\end{figure} 

To obtain the many-body phase diagram for the interacting BEC it is no longer possible to apply the Bloch ansatz as in the previous section. Therefore, obtaining the many-body phase diagram is numerically challenging because it requires a very good momentum space resolution along the $y$ direction (the direction along which the spin-orbit coupling occurs). To this end we find the BEC ground state using the self-consistent algorithm by performing an imaginary time evolution of the coupled Gross-Pitaevskii equations,~\eeref{eqn:coupled_GPE_a} and~\eqref{eqn:coupled_GPE_b}, for $x\in\left[0,\lambda\right]$ and $y\in\left[0,80 \lambda\right]$ with periodic boundary conditions. This implies a quasi-momentum resolution $\Delta q_y = 2\pi\hbar/80=0.08\hbar$ in the $y$ direction, which allows us to sufficiently resolve the two minima in the energy dispersion. We introduce the dimensionless parameter $\gamma=(g-g_{\uparrow\downarrow})/(g+g_{\uparrow\downarrow})$, which tunes the relative strength of intra- and inter-species interactions. In fact, the specific choice of $\gamma$ has no significant effects on the fundamental physics presented in this work and we, therefore, fix it to $\gamma=0.8$.

The phase diagram in~\fref{fig:MB_phase_diag}(a) exhibits three phases. For certain parameter regimes the HOM and the SH phases, which were already identified in the single particle regime, remain the ground states in the interacting case as indicated by the black-white hatched (HOM) and red (SH) regions in~\fref{fig:MB_phase_diag}(a). Note that the two-body interactions also shift the superradiant threshold to larger pump-strength values [green curve in Figs.~\ref{fig:MB_phase_diag}(a) and (b)]. In the SH phase the spin imbalance is always non-zero $\langle\sigma_z\rangle\neq 0$ (red color in the phase diagram). However, the two-body interactions give rise to an additional state---the supersolid spin-density-wave (SSDW) phase. In contrast to the HOM and the SH phase, the spin imbalance vanishes, $\langle\sigma_z\rangle=0$, in this phase as indicated by the blue color code in~\fref{fig:MB_phase_diag}(a). The reason for the zero spin imbalance is that due to the interactions the atoms condense in an equal superposition of the two minima in momentum space to minimize the total energy. This underlying mechanism leading to this additional phase is reminiscent of the mechanism resulting in a stripe phase in interacting spin-orbit-coupled BECs in free space~\cite{martone_anisotropic_2012,li_stripe_2017}.  However, the dynamic coupling of the BEC to the cavity leads to substantially different ground state properties as we will show in the following.

We note that the many-body phase diagram can also be explored non-destructively via the cavity field, as can be seen from~\fref{fig:alpha_PD}.  The phase transition between the HOM and the SH and SSDW phases is always second order [see~\ref{fig:alpha_PD}(c)]. However, at the SH to SSDW transition the absolute value of the cavity fields jumps [see~\ref{fig:alpha_PD}(b)], indicating a first-order phase transition between the SH and the SSDW states. 

\section{Ground states}\label{sec:groundstates}
\begin{figure}
\centering
\includegraphics[width=0.33\textwidth]{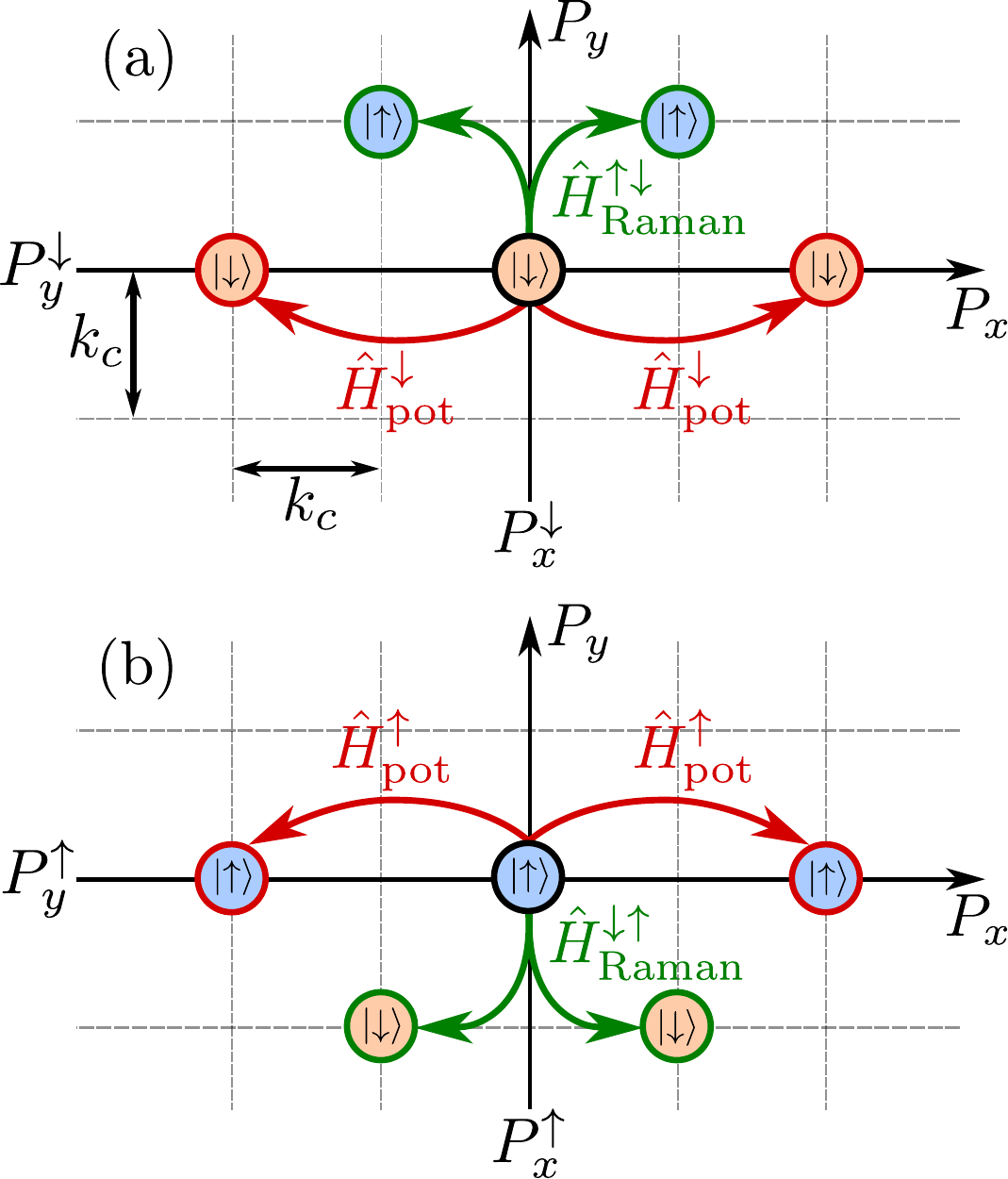}
\caption{Graphical representation of the different terms in the single-particle Hamiltonian in~\eref{eqn:Ham_MB_mom_single_terms}. The red arrows indicate processes governed by the potential Hamiltonians $\hat{H}_\mathrm{pot}^\tau$ and the green arrows indicate Raman processes described by $\hat{H}_\mathrm{Raman}^{\uparrow\downarrow}$ and $\hat{H}_\mathrm{Raman}^{\downarrow\uparrow}$. Note that due to the spin-orbit coupling in general $P_y^{\uparrow}$ differs from $P_y^\downarrow$.}
\label{fig:MB_mom_comp_sketch}
\end{figure}
We now turn to the presentation of typical examples of the various phases discussed above. To understand the ground state properties and underlying mechanisms resulting in different states, we re-write the many-body Hamiltonian in momentum space. It is obtained from~\eeref{eqn:Ham_MB_x}  and~\eqref{eqn:Ham_MB_2_body_int} by writing the field operators as
\begin{equation}
\hat{\Psi}_\tau(\mathbf{r})=\frac{1}{\sqrt{V}}\sum_{\mathbf{P}^\tau}\hat{c}_{\mathbf{P}^\tau,\tau} e^{i\mathbf{P}^\tau\cdot\mathbf{r}},
\label{eqn:field_op_mom}
\end{equation}
where $\hat{c}_{\mathbf{P}^\tau,\tau}$ ($\hat{c}^\dagger_{\mathbf{P}^\tau,\tau}$) is a bosonic annihilation (creation) operator destroying (creating) a particle with spin $\tau\in\{\uparrow,\downarrow\}$ and \emph{kinetic} momentum $\mathbf{P}^\tau$. Note that due to spin-orbit coupling the kinetic momenta $\mathbf{P}^\uparrow$ and $\mathbf{P}^\downarrow$ are not equal in general; \cf~\eref{eqn:physical_mom}.

The many-body Hamiltonian in momentum space is recast as 
$\hat{H}=\hat{H}_0+ \hat{H}_\mathrm{int}$,
with the first term $\hat{H}_0$ describing the non-interacting part,
\begin{align}
&\hat{H}_0=
\sum_{\tau\in\{\uparrow,\downarrow\}}\bigg[\hat{H}_\mathrm{kin}^\tau + \hat{H}_\mathrm{pot}^\tau \bigg] + \hat{H}_\mathrm{Raman}^{\uparrow\downarrow} + \hat{H}_\mathrm{Raman}^{\downarrow\uparrow} + \hat{H}_\mathrm{cav},
\label{eqn:Ham_MB_mom}
\end{align}
where
\begin{align}
&\hat{H}_\mathrm{kin}^\tau=\sum_{\mathbf{P}^\tau}\frac{\hbar^2 |\mathbf{P}^\tau|^2}{2m}\hat{c}_{\mathbf{P}^\tau,\tau}^\dagger \hat{c}_{\mathbf{P}^\tau,\tau}, \nonumber\\
%
&\hat{H}_\mathrm{pot}^\tau= \frac{\hbar U_0}{4}\hat{a}^\dagger\hat{a}
 \sum_{\mathbf{P}^\tau}\sum_{s=\pm 1} \hat{c}^\dagger_{[\mathbf{P}^\tau+s 2\hbar k_c\mathbf{e}_x],\tau} c_{\mathbf{P}^\tau,\tau}, \nonumber\\
%
%
&\hat{H}_\mathrm{Raman}^{\uparrow\downarrow}=\frac{\hbar \eta_0}{2}\left(\hat{a}^\dagger+\hat{a}\right)
\sum_{\mathbf{P}^\downarrow}\sum_{s=\pm 1}
\hat{c}^\dagger_{[\mathbf{P}^\uparrow+\hbar k_c(s\mathbf{e}_x+\mathbf{e}_y)],\uparrow}\hat{c}_{\mathbf{P}^\downarrow,\downarrow}, \nonumber\\
%
&\hat{H}_\mathrm{Raman}^{\downarrow\uparrow}=\frac{\hbar \eta_0}{2}\left(\hat{a}^\dagger+\hat{a}\right)
\sum_{\mathbf{P}^\uparrow}\sum_{s=\pm 1}
\hat{c}^\dagger_{[\mathbf{P}^\downarrow+\hbar k_c(s\mathbf{e}_x-\mathbf{e}_y)],\downarrow}\hat{c}_{\mathbf{P}^\uparrow,\uparrow}, \nonumber\\
&\hat{H}_\mathrm{cav} = \left(\frac{\hbar U_0}{2}-\Delta_c\right) \hat{a}^\dagger\hat{a},
\label{eqn:Ham_MB_mom_single_terms}
\end{align}
and $\mathbf{e}_{x,y}$ are the unit vectors in the $x$ and $y$ directions. $\hat{H}_\mathrm{kin}^\tau$ takes into account the kinetic energy of the two components, the second line $\hat{H}_\mathrm{pot}^\tau$ describes the cavity potential, and the Raman processes are taken into account by $\hat{H}_\mathrm{Raman}^{\uparrow\downarrow}$ and $\hat{H}_\mathrm{Raman}^{\downarrow\uparrow}$. The last term  $\hat{H}_\mathrm{cav}$ is the the cavity Hamiltonian. The kinetic energy Hamiltonian in~\eref{eqn:Ham_MB_mom_single_terms} does not explicitly depend on the cavity mode $\hat{a}$ whereas the other terms all contain the cavity mode operator $\hat{a}$. This again exhibits the dynamic nature of spin-orbit coupling. Once the cavity mode is populated all terms in~\eref{eqn:Ham_MB_mom_single_terms} contribute and spin-orbit coupling sets in. 

Figure~\ref{fig:MB_mom_comp_sketch} shows a graphical interpretation of the different terms contained in the Hamiltonian $H_0$ for both spin components. The interaction of atoms and cavity photons results in a $\pm 2\hbar k_c$ momentum exchange along the cavity axis (i.e., the $x$ axis). This results in the population of the corresponding two additional momentum states indicated in red in~\fref{fig:MB_mom_comp_sketch}. If a photon is scattered from the pump lasers into the cavity (\cf $H_\mathrm{Raman}^{\downarrow\uparrow}$) a momentum transfer of $\pm \hbar k_c$ is imposed in both $x$- and $y$ directions. This processes (green arrows in~\fref{fig:MB_mom_comp_sketch}) results in the population of two additional momentum states indicated in green in~\fref{fig:MB_mom_comp_sketch}. These terms also show that photons can only be scattered into the cavity via a spin-flip process $\ket{\uparrow}\leftrightarrow\ket{\downarrow}$, but not via the BEC density.

The two-body interaction Hamiltonian for two particles with two momenta $\mathbf{P}_{1,2}^\tau$ exchanging a momentum $\mathbf{K}$ takes the form
\begin{align}
\hat{H}_\mathrm{int} &= \frac{g}{2}\sum_{\tau\in\{\uparrow,\downarrow\}}\sum_{\mathbf{P}_1^\tau,\mathbf{P}_2^\tau,\mathbf{K}}\hat{c}^\dagger_{\mathbf{P}_2^\tau-\mathbf{K},\tau}\hat{c}^\dagger_{\mathbf{P}_1^\tau+\mathbf{K},\tau}\hat{c}_{\mathbf{P}_1^\tau,\tau}\hat{c}_{\mathbf{P}_2^\tau,\tau}\nonumber\\
&+g_{\uparrow\downarrow}\sum_{\mathbf{P}_1^\uparrow,\mathbf{P}_2^\downarrow,\mathbf{K}}
\hat{c}^\dagger_{\mathbf{P}_2^\downarrow-\mathbf{K},\downarrow}\hat{c}^\dagger_{\mathbf{P}_1^\uparrow
+\mathbf{K},\uparrow}\hat{c}_{\mathbf{P}_1^\uparrow,\uparrow}\hat{c}_{\mathbf{P}_2^\downarrow,\downarrow}.
\label{eqn:Ham_MB_2-body_int_mom}
\end{align}
The Hamiltonians~\eqref{eqn:Ham_MB_mom_single_terms} and~\eqref{eqn:Ham_MB_2-body_int_mom} allow us to get an intuitive picture that which momentum states can be populated via photon scattering processes. However, this does not imply that all these momentum components are ultimately populated. As we will show in the following the intuitive argument based on the analysis of the different terms of the many-body Hamiltonian in momentum space is in good agreement with the numerically obtained ground-state momentum distributions.

\subsection{Spin-Helix Phase}
\label{sec:SH-phase}

\begin{figure}[t!]
\centering
\includegraphics[width=0.5\textwidth]{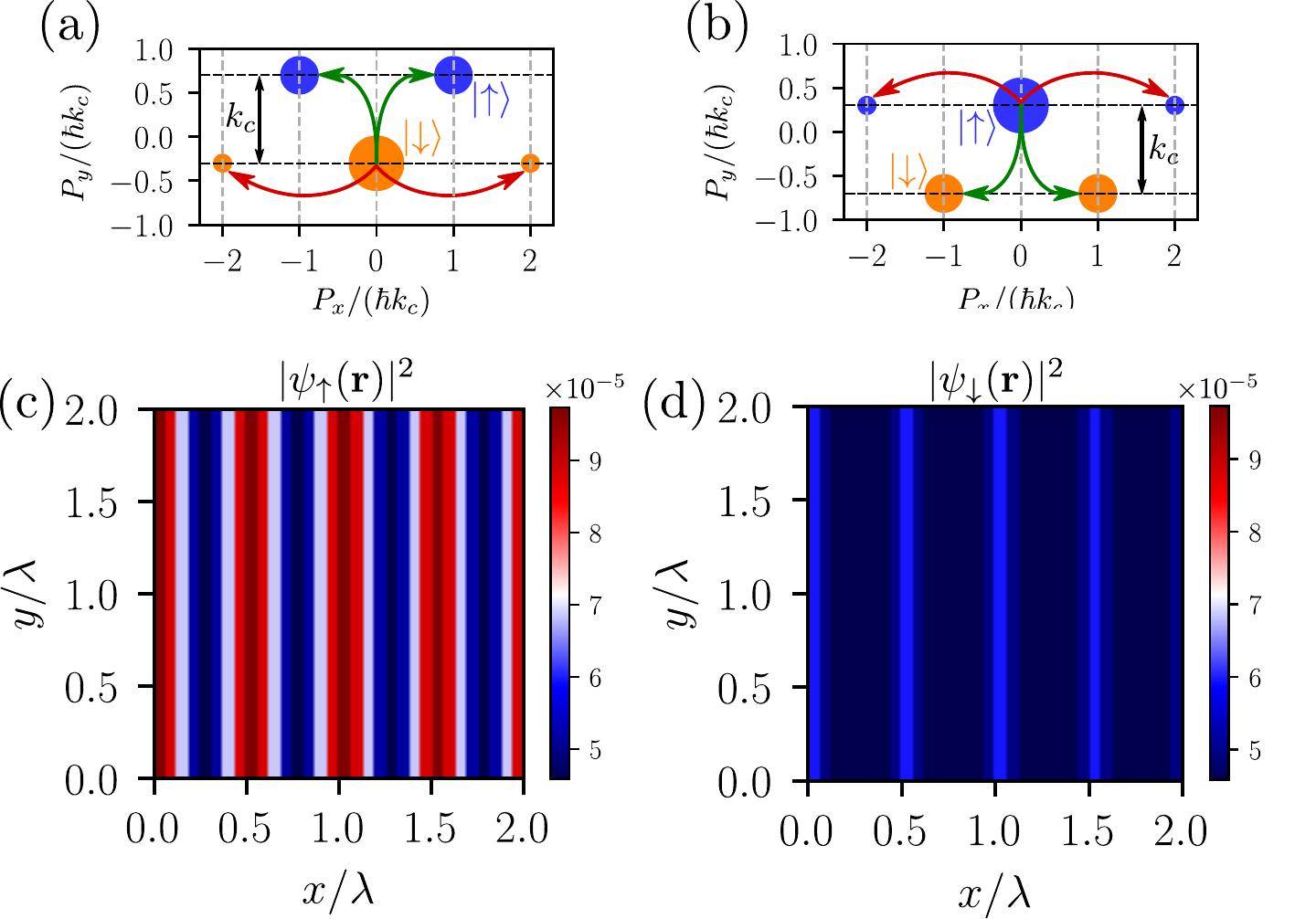} 
\caption{Typical ground state for the spin-helix (SH) phase  for $\eta_0=4.0\omega_\mathrm{rec}$ and $gn=0\omega_\mathrm{rec}$. (a) and (b) show the two possible momentum space configurations. Orange circles correspond to the psuedospin-down component $\ket{\downarrow}$ and blue circles to the pseudospin-up component $\ket{\uparrow}$. The size of the circles visualizes the population of the corresponding momentum state. The panels (c) and (d) show the position-space density distribution obtained from the momentum space distribution shown in panel (b) for the two pseudospin components. The particle number in the $\ket{\uparrow}$-state is $N_\uparrow=0.71 N$ and in the $\ket{\downarrow}$-state $N_\downarrow=0.29 N$. The other parameters are the same as in~\fref{fig:MB_phase_diag}.}
\label{fig:spin_helix}
\end{figure}

\begin{figure}
\centering
\includegraphics[width=0.42\textwidth]{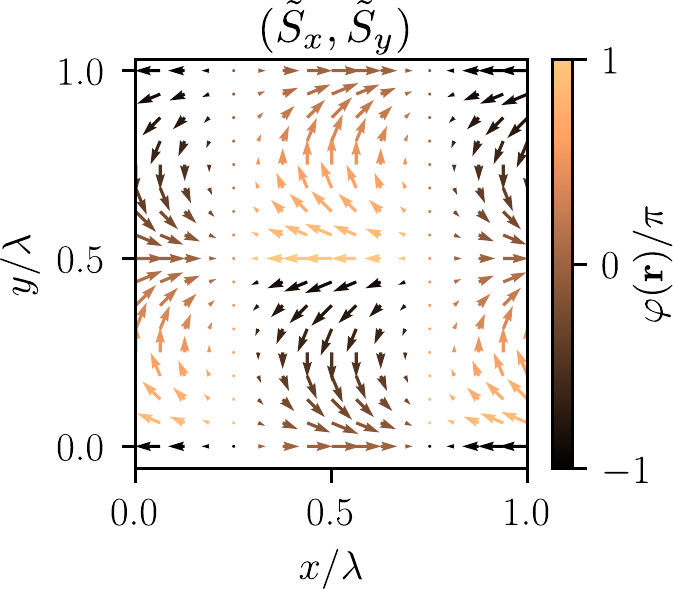}
\caption{Exemplary spin texture for the spin-helix (SH) phase. It is visualized via the projection of the normalized spin vector $\tilde{\mathbf{S}}(\mathbf{r})$ in the $\tilde{S}_x$-$\tilde{S}_y$ plane (d). The color indicates the spin angle $\varphi(\mathbf{r})=\tan^{-1}(\tilde{S}_y/\tilde{S}_z)$. The spin performs a full spiral (i.e., a $2\pi$ rotation) along the $y$ direction in the $x$-$y$ plane. The parameters are the same as in~\fref{fig:spin_helix}.}
\label{fig:spin_helix_spin}
\end{figure}
In the SH phase the particles condense in one of the two minima of the dispersion relation. Hence, one of the two configurations shown in~\fref{fig:MB_mom_comp_sketch} is realized randomly. In~\ffref{fig:spin_helix}(a) and (b) the two numerically obtained momentum state distributions for the two possible ground states are shown. They perfectly coincide with the intuitive picture based on the Hamiltonian in momentum space (\cf~\fref{fig:MB_mom_comp_sketch}). Note that due to the Raman processes some particles are transferred from $\ket{\downarrow} \rightarrow \ket{\uparrow}$ or $\ket{\uparrow} \rightarrow \ket{\downarrow}$ but the spin imbalance $\langle\sigma_z\rangle$ is still always non-zero because the population of the momentum component corresponding to the initial condensate always remains bigger than the two momentum states populated via the Raman processes; see~\fref{fig:MB_phase_diag}. The $y$ component of the kinetic momentum at which condensation takes place is fixed by the value of the quasi-momentum corresponding to the energy-dispersion minimum (or minima). Following from the relation given in~\eref{eqn:physical_mom} the $y$ component of the kinetic momentum for a given quasi-momentum $q_y$ for each spin component can be calculated via $P_y^\uparrow=q_y + \hbar k_c 2$ and $P_y^\downarrow=q_y - \hbar k_c 2$.

Typical real-space density distributions of the two pseudospin components in the SH phase are shown in~\ffref{fig:spin_helix}(c) and (d). Due to the emergent cavity potential a $\lambda/2$-periodic density modulation along the $x$ axis is formed, but no density modulation along the $y$ axis occurs. Therefore, the SH phase does not break the continuous symmetry of the Hamiltonian along the $y$ direction. The spin texture forms a helix in the $y$ direction as illustrated in \fref{fig:spin_helix_spin}, where we plot the projection of the normalized total spin 
\begin{align}
\tilde{\mathbf{S}}(\mathbf{r})=\mathbf{S}(\mathbf{r})/\sqrt{S_x^2(\mathbf{r})+S_y^2(\mathbf{r})+S_z^2(\mathbf{r})}, 
\end{align}
on the $\tilde{S}_x$-$\tilde{S}_y$ plane. Here $\mathbf{S}(\mathbf{r})=\langle \mathbf{\hat{S}}(\mathbf{r})\rangle$ is the local mean-field spin vector, with $\hat{\mathbf{S}}(\mathbf{r})=\hat{\Psi}^\dagger(\mathbf{r})\pmb{\sigma}\hat{\Psi}(\mathbf{r})$ ($\pmb{\sigma}$ is the vector of the Pauli-matrices). The color coding  in \fref{fig:spin_helix_spin} corresponds to the spin angle $\varphi(\mathbf{r})=\tan^{-1}(S_y/S_z)$. It exhibits the $Z_2$-symmetry breaking in the spin-domain~\cite{mivehvar_cavity-quantum-electrodynamical_2019}. This phase is intimately related to the plane-wave phase in 1D spin-orbit-coupled BECs in free space. Note that the SH phase can also be realized in the non-degenerate case ($\delta\neq 0$) where the energies of the two pseudospin states $\ket{\downarrow}$ and $\ket{\uparrow}$ do not coincide. This is the regime where previous theoretical~\cite{mivehvar_cavity-quantum-electrodynamical_2019} and experimental works~\cite{kroeze_dynamical_2019} focused on. Of course, in this case only one of the two states shown in~\ffref{fig:spin_helix}(a) and (b) can be realized depending on the sign of $\delta$.

\subsection{Supersolid Spin-Density-Wave Phase}

\begin{figure*}[t!]
\centering
\includegraphics[width=0.75\textwidth]{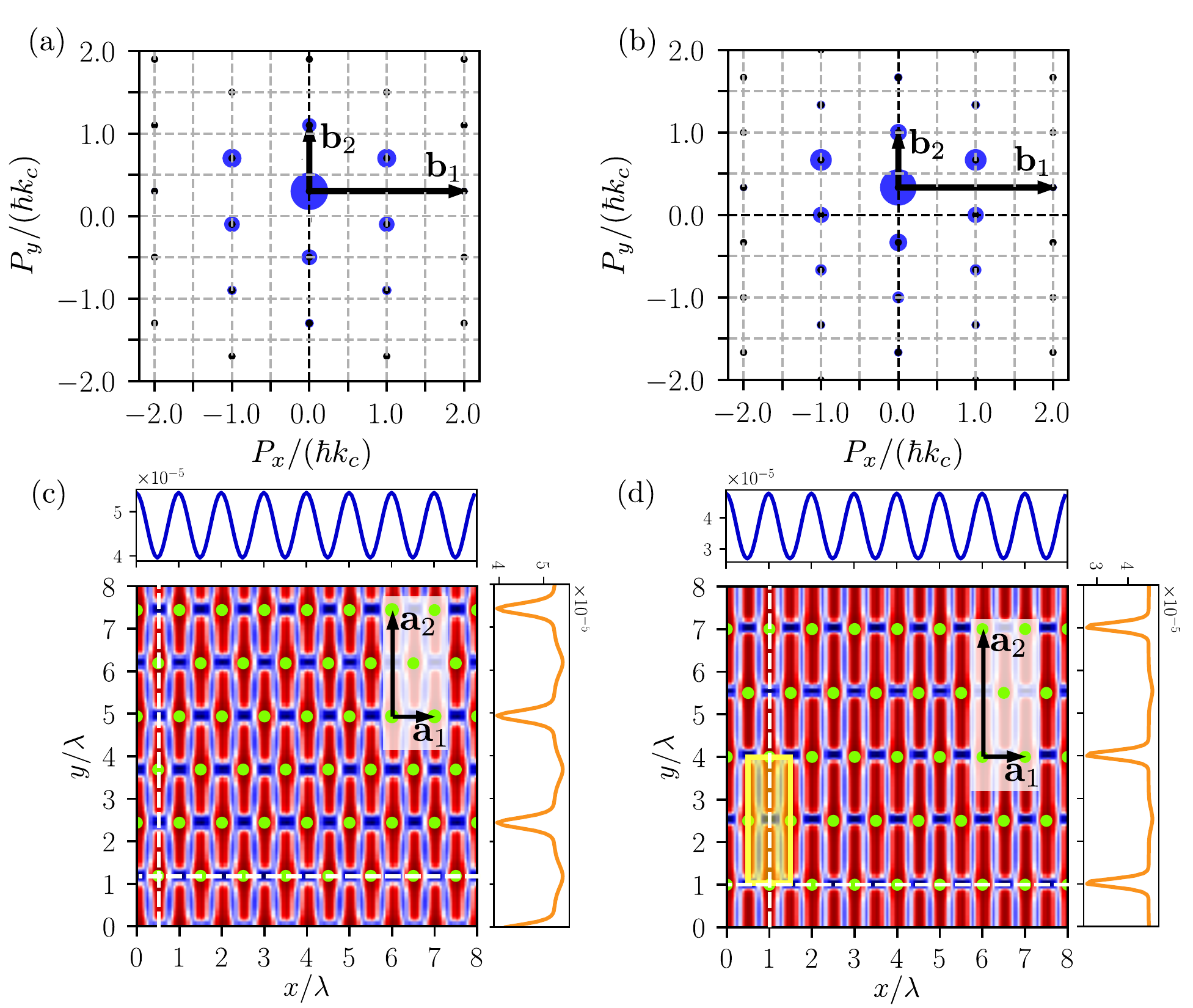} 
\caption{Exemplary momentum state populations (a)--(b) and the corresponding density distributions (c)--(d) in the SSDW for two different pump strengths $\eta_0 = 4.0\omega_\mathrm{rec}$ (left), $4.5\omega_\mathrm{rec}$ (right) and $gn=10\omega_\mathrm{rec}$ for the pseudospin $\ket{\uparrow}$ component. The green dots mark the density maxima. The blue (orange) curves at the boundary of the 2D plot show cuts through the density distributions along the white dashed lines on the $x$ ($y$) direction. The density distribution forms a centered orthorhombic lattice spanned by the vectors $\mathbf{a}_1$ and $\mathbf{a}_2$. For increasing pump strengths the periodicity in $y$ direction becomes larger (smaller) in position (momentum) space due to the spin-orbit coupling [\cf~\fref{fig:PD_non-int}(b)]. The vectors $\mathbf{b}_i$ are the reciprocal lattice vectors determined by $\mathbf{a}_i$. An exemplary spin texture for the region indicated in yellow in panel (d) is shown in~\fref{fig:supersolid_spin_text}. The other parameters are the same as in~\fref{fig:MB_phase_diag}.}
\label{fig:ex_supersolid_spin_wave}
\end{figure*}

The two-body interaction Hamiltonian~\eqref{eqn:Ham_MB_2-body_int_mom} adds two additional terms to the many-body Hamiltonian, which give rise to interaction induced pattern formation and spontaneous symmetry breaking in the $y$ direction. The underlying process for this pattern formation is the coherent multi-mode mixing of coherent BEC momentum components. This effect is a very fundamental property of Bose-Einstein condensates, often also referred to as ``matter-wave interference". This genuine quantum effect, which is directly related to the off-diagonal long-range order,~\ie the coherence of the BEC, ultimately leads to the breaking of the continuous symmetry and the emergence of an orthorhombic centered rectangular-lattice density pattern in some parameter regimes; cf.~\fref{fig:ex_supersolid_spin_wave}.

\begin{figure}[t!]
\includegraphics[width=0.49\textwidth]{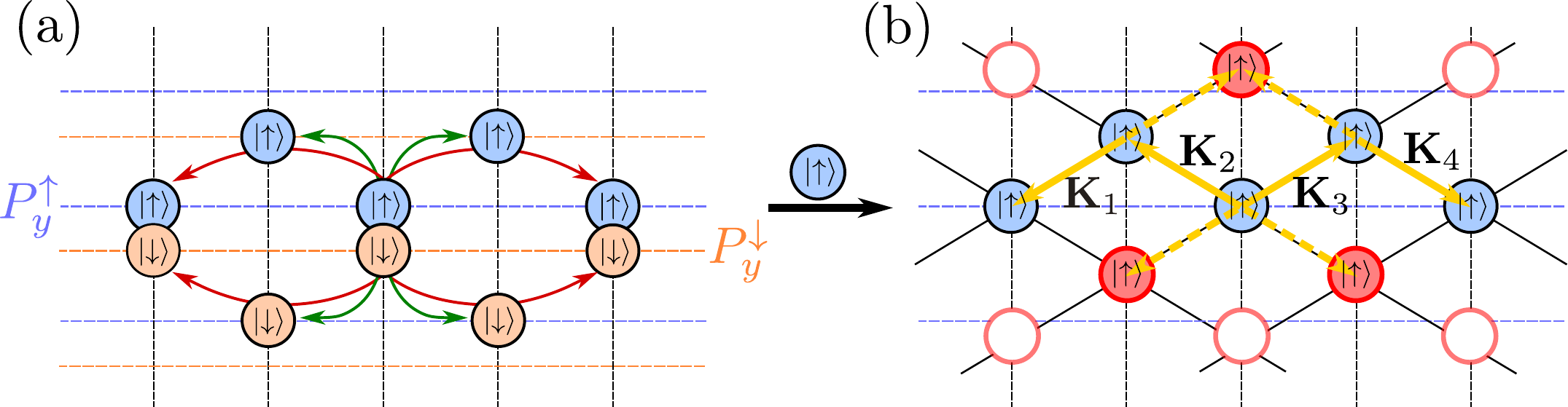} 
\caption{Graphical illustration of the effects leading to the SSDW phase. (a) The system condenses in a superposition of the two possible cases shown in~\fref{fig:MB_mom_comp_sketch}. (b) Coherent multi-mode mixing of the different momenta due to the two-body intra-species interactions $\propto g$ results in the population of additional momentum states (red circles). This leads to a periodic density modulation with an orthorhombic centered rectangular-lattice pattern.}
\label{fig:MB_sketch_interactions}
\end{figure}

In the previous section we discussed two possible momentum space configurations for the SH phase. In the SSDW phase the particles condense into an equal superposition of the two cases shown in Fig.~\ref{fig:MB_mom_comp_sketch} due to the inter-species interactions. Based on the intuitive discussion above this results in a momentum distribution as sketched in~\fref{fig:MB_sketch_interactions}(a). Note that this also implies a vanishing spin imbalance $\langle \sigma_z \rangle=0$ as it was already suggested by the many-body phase diagram in \fref{fig:MB_phase_diag}. This feature is reminiscent of the mechanism leading to a periodic stripe phase in 1D spin-orbit-coupled BECs in free space~\cite{lin_spinorbit-coupled_2011, martone_anisotropic_2012}. However, the presence of the cavity results in the population of different momentum states compared to the free space case, which ultimately results in a 2D orthorhombic centered rectangular-lattice density structure, as will be discussed in the following. For the sake of simplicity, we only focus on the $\ket{\uparrow}$ component. The same arguments apply to the $\ket{\downarrow}$ component as well. The only difference in this case is that the momentum space distribution is mirrored around the zero line of the $P_x$-axis. However this results in exactly the same density distributions for the pseudospin $\ket{\downarrow}$ component.

The condensation in the equal superposition state results in the population of multiple momentum states of the same pseudospin component; see blue and orange circles in~\fref{fig:MB_sketch_interactions}(a). In particular momentum states with different momenta in $P_y$ direction are now populated for the same spin component, which was not the case in the SH phase. These  momentum components can be coherently mixed via the intra-species interaction Hamiltonian. This coherent multi-mode mixing process is described by the first term of the Hamiltonian~\eqref{eqn:Ham_MB_2-body_int_mom} and it results in the population of additional momentum states as shown by red circles in~\fref{fig:MB_sketch_interactions}(b). For example, the momentum exchange during the (repulsive) s-wave scattering process between particles in two momentum states with a momentum difference $\hbar|\mathbf{K}_i|$ [solid yellow arrows in Fig.~\ref{fig:MB_sketch_interactions}(b)] results in the creation of two particles with momenta $\pm \hbar\mathbf{K}_i$ as it can be directly seen from~\eref{eqn:Ham_MB_2-body_int_mom}. This coherent collissional coupling gives rise to the circles with red filling in Fig.~\ref{fig:MB_sketch_interactions}(b). Higher order couplings result in even additional momentum states indicated by the white circles with red border in~\fref{fig:MB_sketch_interactions}. This mechanism can also be understood as four-wave mixing of matter waves~\cite{deng_four-wave_1999, goldstein_quantum_1999, trippenbach_theory_2000, hung_four-wave_2019}.

Typical momentum and density distributions for the SSDW phase are shown in~\fref{fig:ex_supersolid_spin_wave}. The momentum space distributions resemble the intuitive picture depicted in~\fref{fig:MB_sketch_interactions}(b). Indeed additional momentum states are populated in the ground state due to the two-body interactions; see~\ffref{fig:ex_supersolid_spin_wave}(a) and (b). The corresponding density pattern breaks the continuous symmetry in the $y$ direction and forms a centered orthorhombic lattice where the unit cell is spanned by the lattice vectors $\mathbf{a}_1$ and $\mathbf{a}_2$. The corresponding reciprocal lattice vectors are defined as $\mathbf{b}_1=2\pi[\mathbf{a}_2\times\mathbf{a}_3]/[\mathbf{a}_1\cdot(\mathbf{a}_2\times\mathbf{a}_3)]$, $\mathbf{b}_2=2\pi[\mathbf{a}_3\times\mathbf{a}_1]/[\mathbf{a}_1\cdot(\mathbf{a}_2\times\mathbf{a}_3)]$ and $\mathbf{b}_3=2\pi[\mathbf{a}_1\times\mathbf{a}_2]/[\mathbf{a}_1\cdot(\mathbf{a}_2\times\mathbf{a}_3)]$ with $\mathbf{a}_3=\mathbf{e}_z\coloneqq(0,0,1)^\top$. 

The absolute values $|\mathbf{a}_1|$ and $|\mathbf{b}_1|$ are set by the cavity potential (i.e., the cavity wave length $\lambda_c$). However, due to the different minimum positions of the energy dispersion (\cf \fref{fig:dispersion}), the periodicity of the density distribution in the $y$ direction changes and $|\mathbf{a}_2|$ ($|\mathbf{b}_2|$) increases (decreases) for growing pump strengths. Note that the periodicity in the $y$ direction is solely governed by the quasi-momenta corresponding to the minima of the energy dispersion (see~\fref{fig:PD_non-int}). This is fundamentally different from other self-organization phenomena which solely rely on the built-up of a superradiant optical lattice. In this case the lattice spacing in the $x$ \emph{and} $y$ direction is in general a multiple of the cavity resonance wavelength $\lambda_c$. However, this is not the case in the dynamic spin-orbit-induced many-body phase discussed in this section. Since the emergent density distribution shown in~\ffref{fig:ex_supersolid_spin_wave}(c) and (d) spontaneously breaks the continuous symmetry in the $y$ direction, the system has supersolid properties~\cite{boninsegni_colloquium_2012}. The formation of the SSDW phase has certain analogies to the steering of matter-wave superradiance with an optical cavity~\cite{kesler_steering_2014}. However, due to the presence of spin-orbit coupling the periodicity in the $y$ direction is no longer solely fixed by the cavity resonance wavelength.

The SSDW phase emerges due to two fundamental properties of the system: coherent  scattering of photons \emph{and} coherent multi-mode mixing of the BEC momentum components. The former results in cavity-induced spin-orbit coupling and the population of the momentum states shown in~\fref{fig:MB_sketch_interactions}(a). The latter couples these different momentum states due to two-body interactions, which results in non-trivial density distribution; see~\ffref{fig:MB_sketch_interactions}(b) and~\ref{fig:ex_supersolid_spin_wave}. Consequently, the coherence of the condensate plays a crucial role in the formation of the periodic density distribution. In this respect the studied system differs substantially from other systems exhibiting self-ordering in optical resonators~\cite{ritsch_cold_2013}. In most systems it is the coherent cavity field forming an emergent optical lattice, which results in a self-organized periodic density pattern, hence the coherence of the BEC and two-body interaction do not play a major role~\cite{brennecke_real-time_2013}. Therefore, cold thermal gases can also exhibit self-ordering if they interact with a coherent cavity field. This, however, is no longer true in the case of the SSDW phase presented here. For the periodic density formation discussed here, both processes---the coherent scattering of photons into the cavity \emph{and} the coherence of the BEC momentum components as well as the two-body interactions resulting in multi-mode mixing---are crucial. A related pattern formation process in a BEC via multi-mode mixing of different momentum states was also experimentally observed recently in a driven BEC with modulated interaction strengths~\cite{zhang_pattern_2020}. The formation of the density pattern in our model also shares some aspects of the formation of supersolid droplets in dipolar BECs where the combination of long-range dipolar interactions and local repulsive interactions results in stable droplet solutions~\cite{chomaz_long-lived_2019, bottcher_transient_2019,natale_excitation_2019, tanzi_observation_2019}.

\begin{figure}
\centering
\includegraphics[width=0.46\textwidth]{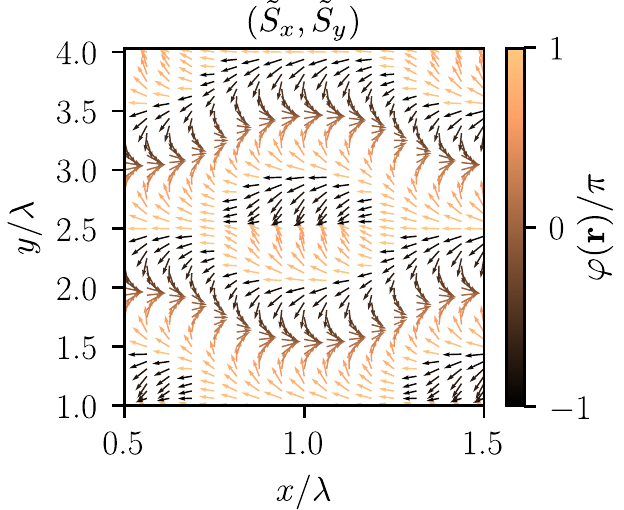} 
\caption{Projection of the normalized spin vector $\tilde{\mathbf{S}}(\mathbf{r})$ in the $\tilde{S}_x$-$\tilde{S}_y$ plane for the region indicated in yellow in~\fref{fig:ex_supersolid_spin_wave}(d) in the SSDW phase for $\eta_0=4.5\omega_\mathrm{rec}$. The other parameters are the same as in~\fref{fig:MB_phase_diag}.}
\label{fig:supersolid_spin_text}
\end{figure}

The additional density modulation along the $y$ direction also changes the spin texture by changing the length of the spin vectors $|\tilde{\mathbf{S}}(\mathbf{r})|$ locally; see~\fref{fig:supersolid_spin_text}. In other respects, the spin textures still exhibit the spiral nature in $y$ direction. However, at the regions around the density minima the spin spiral is altered significantly and it exhibits a jump in the spin direction $\propto 2\pi$. This could be due to the interplay between cavity-mediated global spin interactions~\cite{mivehvar_cavity-quantum-electrodynamical_2019} and two-body collision-induced local spin interactions, which its investigation goes beyond the scope of the present publication and will be considered elsewhere.

\section{Conclusions and Outlook}\label{sec:conclusion}

In conclusion, the theoretical analysis of the single particle and many-body properties of a planar spinor BEC coupled to a single mode of a standing-wave resonator reveals a very rich phase diagram. The presence of the cavity alters the physics substantially compared to free space spinor BECs. On the single particle level the dynamic cavity potential modifies the energy dispersion such that the region where the dispersion exhibits two minima with different quasi-momenta extends over a much larger parameter regime compared to free space case. For the interacting BEC the additional momentum components populated due to coherent photon scattering into the cavity forms the basis of an additional transverse periodic density wave resulting in a 2D centered orthorhombic lattice for certain parameters. In contrast to conventional cavity-induced self-ordering~\cite{ritsch_cold_2013}, this pattern formation is due to the combination of coherent photon scattering into the cavity mode \emph{and} coherent momentum mixing via local two-body collisional interactions. As collisions here do not work against density-wave order but are essential to create diagonal order, our findings can lead to a new paradigm in the self-ordering of BECs in resonators, where the off-diagonal long-range order and interactions of the BEC play a crucial role.

The experimental geometry to study the presented phenomena is up to some minor modifications in the laser geometries already realized in several labs. Still the experimental realization requires good control of the interaction strength, which is a challenge to be overcome in order to realize the predicted SSDW phase. Nevertheless, we believe that the predicted phase diagram can be studied in state-of-the-art experiments. In general, even more complex pattern formation could be observed by taking into account more resonator modes, more atomic levels, and/or different pump-laser geometries. The studied setup also exhibits the potential of cavity-QED systems for implementing dynamical artificial gauge fields for neutral atoms.\\[2ex]

\emph{Acknowledgments.} We would like to thank Elvia Colella and Karol Gietka for fruitful discussions.  This work was supported by the international Joint Project No.\ I3964-N27 of the Austrian Science Fund (FWF) and the National Agency for Research (ANR) of France. F.\,M.\ also acknowledges funding by the FWF Lise-Meitner Fellowship M2438-NBL.

The numerical simulations were performed with the open-source framework \texttt{QuantumOptics.jl}~\cite{kramer_quantumopticsjl_2018}. The authors would like to thank David Plankensteiner for technical support on using the framework.

%

\end{document}